\documentclass[11pt,graphicx]{article} \textwidth 16cm \textheight 22cm
\usepackage{graphics}

\title{Non-spectator Contributions To  The Lifetime of $\Lambda_{b}$ }
\author{
Chao-Hsi Chang$^{1,2}$, \ \   Peng Guo$^3$, \ \   Xue-Qian Li$^3$ \  and \   Guo-Li Wang$^3$ \\
$^1$  CCAST (World Laboratory), P.O. Box 8730, Beijing 100080,
China\\
$^2$  Institute of Theoretical Physics, Academia Sinica, P.O. Box
2735, Beijing 100080, China \\
 $^3$  Department of Physics, Nankai University,
Tianjin 300071,
China \\
 }

\begin{document}
\maketitle

\begin{abstract}
In this work, we evaluate the contributions  of non-spectator
effects to the lifetimes of $\Lambda_b$ and B-mesons. Based on the
well-established models and within a reasonable range of the
concerned parameters, the contributions can reduce  the lifetime
of $\Lambda_b$  by $7 \sim 8\%$ compared to that of B-mesons which
are not significantly affected. This might   partly explain the
measured ratio $\tau(\Lambda_{b})/\tau(B^{0})=0.79 $ \cite{Data},
which has been a long-standing discrepancy between theory and
experimental data.

\end{abstract}

\section{Introduction}

The heavy-flavor world provides us an opportunity to get insight
into the fundamental physics. The most intriguing  problem at the
present stage of the theoretical studies on hadron properties and
the processes of relatively low energies is lack of solid
knowledge on non-perturbative QCD. The non-perturbative QCD
effects which govern the hadronic transition matrix elements, are
entangled  with the hard subprocesses.

For the processes where some heavy flavors (b and/or c) are
involved, by the heavy quark effective theory(HQET)
\cite{Isgur,Georgi}, the short and long-distance QCD effects are
separated in a systematic way. Beneke et al.\cite{Beneke}
demonstrated how to correctly apply the factorization procedure to
the processes where heavy mesons transit into light ones.

On other side, there are still some unsolved problems in the heavy
flavor physics. A protrudent problem is the lifetime of
$\Lambda_{b}$. The present data for the ratios are \cite{Data}
 \begin{eqnarray}
 \frac{\tau(B^{-})}{\tau(B_{d})}&=&1.06\pm0.04,      \\
 \frac{\tau(\Lambda_{b})}{\tau(B_{d})}&=&0.79\pm0.06.
 \end{eqnarray}

In the traditional spectator scenario,  if there are no strong
decay channels the weak decays of b-hadrons  as well as their
lifetimes  are fully determined by the weak decays of b-quark,
thus  the above ratios must be very close to unity. For D-meson,
the lifetime of $D^{\pm}$ is almost 2.55 times larger than that of
$D^{0}$. This difference is perfectly explained  in the QCD
framework by many authors \cite{Bigi}, where the non-spectator
effects play a crucial role as the Pauli-Interference(PI) strongly
suppresses the total width of $D^{\pm}$ compared to $D^{0}$, but
the effects are not so obvious for $B^{\pm}$ and $B^{0}$ due to
the $\Lambda_{QCD}/m_{b}$ suppression. We would think that the
lifetimes of $D^{\pm}$, $D^0$ and $B^{\pm}$, $B^0$ are well
understood, but for the $\tau(\Lambda_b)$ so far there is not a
satisfactory answer to the puzzle yet, it is also worth noticing
that this problem is even more serious in the $\Lambda_c$ lifetime
which we will discuss in our coming work.

As one attempts to explain the puzzle of the $\Lambda_b$ lifetime,
it is natural to consider the non-spectator contributions. Even
though the study of the lifetime difference of $B^{\pm}$ and
$B^{0}$ indicates that the non-spectator contribution can only
result in a small effect for B-meson cases because of the
$\Lambda_{QCD}/m_{b}$ suppression, it still is necessary to
investigate how large such effects can be in the b-baryon case.
Namely  before one can claim a new physics or mechanism which
result in the difference, he has to thoroughly explore possible
explanations in the standard QCD framework.

To calculate the lifetime, one only needs to deal with the
inclusive processes   which are relatively simple compared to the
exclusive processes, therefore the results are more reliable.

In this work we consider the contributions from the non-spectator
effects to the lifetime of $\Lambda_b$. At the quark level, the
foundation is the weak effective  Lagrangian \cite{He}. We not
only consider the W-exchange(WE) diagrams and the Pauli
interference(PI) diagrams \footnote{Here PI refers to the
mechanism where a constituent of the final state joins the b-quark
of the initial state at the effective vertex and vice versa.
Namely there exists a crossing of constituent lines (see Figs.1
and 2 for illustration). In the B-meson case, it is indeed a Pauli
interference, but for the $\Lambda_b$ it is only a manifestation
of such mechanisms. }where only two quarks take part in the
reaction with the other one being a spectator, but also account
for the corresponding diagram where all the three valence quarks
are involved. The last mechanism does not exist in the meson case,
and seems to result in a difference between $\tau(\Lambda_{b})$
and $\tau(B^0(B^{\pm}))$.

Moreover we also consider the diquark structure where the two
light quarks constitute a boson-like diquark of color anti-triplet
and it participates in the reaction of WE or PI as a whole
subject. Then we compare the results obtained in the
three-valence-quark picture and the heavy-quark-light-diquark
picture. It could also be a test of the supposed diquark structure
of baryons.

The most difficult task is to evaluate the hadronic transition
matrix elements which are fully governed by the non-perturbative
QCD. Even though we are able to separate the long-distance effects
from the short distance subprocess, we still cannot evaluate the
hadronic matrix elements in a well-established way, but need to
carry out a model-dependent calculation. In this work we adopt the
simplest non-relativistic harmonic oscillator model \cite{Olivier}
for the hadron wavefunctions. To obtain the concerned parameters
in the model, we first evaluate the semi-leptonic decays of
$\Lambda_{b} \rightarrow \Lambda_{c} + e^-+ \bar{\nu}$ and
$B\rightarrow D+e+\bar{\nu}$ which are not contaminated by the
final state interactions. Our results are comparable with  the
works on this aspect in the literatures \cite{Mohanta} and a
recent work by Chakraverty et al \cite{D. Chakraverty}. Then with
the parameters, we calculate the lifetimes of B-mesons and
$\Lambda_b$.

To  evaluate the ratios $\tau(\Lambda_{b}) / \tau(B^{0})$   and
$\tau(B^{\pm}) / \tau(B^{0})$,  we calculate the non-spectator
contributions to the lifetimes of  $B^{0}$ and $B^{\pm}$ in the
same model, namely we achieve the wavefunctions of $B^{0}$ and
$B^{\pm}$  in the two-body harmonic oscillator model. Evaluating
the hadronic matrix elements of $<B \mid {\cal O} \mid B>$ which
appears in the semi-leptonic decay rate of B-mesons, and then by
fitting the data, we gain the model parameters. We hope that since
we evaluate lifetimes of both $\Lambda_{b}$ and B-meson in the
same model, the model-dependence of the theoretically evaluated
ratios can be partly cancelled, and the results would make more
sense.

Our paper is organized as follows. After this introduction we
briefly introduce the non-relativistic harmonic  oscillator model
for $\Lambda_{b}$ in the picture with three valence quarks and as
well as that with one heavy quark and one light diquark, then we
discuss B-mesons in the same model. In section 3, we calculate the
semi-leptonic decays of B-mesons and $\Lambda_{b}$ where  we also
use the two pictures of 3-quark and quark-diquark structures
respectively. In sect.4, we use the model to evaluate the
lifetimes of   B-mesons and $\Lambda_{b}$, then in sect.5, we
present our numerical results. The last section is
devoted to our conclusion and discussion.\\

\section{The wavefunctions of b-hadrons in the non-relativistic harmonic oscillator quark model} \label{sec.1}
\subsection{For the B-mesons }
In the valence quark model, B-mesons contain a heavy quark b and
one light flavor. The spatial coordinates of the two constituents
in B-mesons are denoted by $\mathbf{r}_{b}$ and $\mathbf{r}_{q}$.
It is convenient to introduce the Jacobi coordinates $\mathbf{R}$
and $\mathbf{\rho}$ which are defined as  \\
\begin{eqnarray}
\left\{
\begin{array}{c}
 \mathbf{R}=\frac{ m_{q} \mathbf{r}_{q} +
M_{b}
\mathbf{r}_{b}}{m_{q} +M_{b}} , \\
\mathbf{\rho} = \mathbf{r}_{b} -  \mathbf{r}_{q}.
\end{array}  \right.
\end{eqnarray} \\
The corresponding canonical conjugate momenta $\mathbf{P}$ and
$\mathbf{p}_{\rho}$  are
respectively,\\
\begin{eqnarray}
\left\{
\begin{array}{c}
\mathbf{P}= \mathbf{p}_{b} + \mathbf{p}_{q} , \\
\mathbf{p}_{\rho} = \frac{m_{q} \mathbf{p}_{b} - M_{b}
\mathbf{p}_{q}}{M_{b} + m_{q}} .
\end{array}  \right.
\end{eqnarray} \\
In the non-relativistic harmonic oscillator  model, the normalized
state vector is
written as,\\
\begin{equation}
\mid B_{M}(\mathbf{P},s) > = A_{M} \sum_{color,spin}
C^{s}_{s_{q},s_{b}} \chi_{s_{q},s_{b}} \varphi_{color} \int d^{3}
\mathbf{p}_{\rho} \psi_{B_{M}}(\mathbf{p}_{\rho}) \mid
q(\mathbf{p}_{q},s_{q});\overline{b}(\mathbf{p}_{b},s_{b}) >,
\end{equation}\\
where $\chi$, $\varphi$ and $\psi$ are the flavor-spin, color and
spatial wavefunctions in the momentum space  respectively. $A_{M}$
is a normalization factor, thus $\mid B_{M}(\mathbf{P},s) >$
satisfies the normalization condition
\begin{equation}
<  B_{M}(\mathbf{P},s)   \mid B_{M}(\mathbf{P'},s') > = (2
\pi)^{3} 2 \omega_{P}  \delta^{3} (\mathbf{P} - \mathbf{P'})
\delta_{s,s'}.
\end{equation}
$C^{s}_{s_{q},s_{b}}$ are the Clebsch-Gordon coefficients for
combining two  spin-$\frac{1}{2}$ constituent quarks into a spin-0
meson. The wave function $\psi_{B_{M} }(\mathbf{p}_{\rho}) $ is
the ground state eigenfunction of the harmonic oscillator
hamiltonian and has the form
\begin{equation}
\psi_{B_{M}}(\mathbf{p}_{\rho})= \exp (-
\frac{\mathbf{p}^{2}_{\rho}}{ 2 \alpha^{2}_{M}}),
\end{equation}
where the parameter $\alpha_{M}$ is a free parameter of the model and to be determined
by fitting data of semileptonic decays \cite{Mohanta}.\\

\subsection{For $\Lambda_b$  }

(a) 3-valence-quark model\\

In a non-relativistic quark model (NRQ), $\Lambda_b$ contains a
heavy quark b and two light quarks (u and d). The spatial
coordinates of the three quarks in $\Lambda_b$ are denoted by
$\mathbf{r}_{b}$, $\mathbf{r}_{q_{1}}$ and $\mathbf{r}_{q_{2}}$.
Similar to the meson case, we also introduce the Jacobi
coordinates $\mathbf{R}$,
$\mathbf{\rho}$ and $ \mathbf{\lambda}$ defined as \cite{Bhaduri} \\
\begin{eqnarray}
\left\{
\begin{array}{l}
\mathbf{R}=\frac{ m \mathbf{r}_{q_{1}} + m  \mathbf{r}_{q_{2}} +
M_{b}
\mathbf{r}_{b}}{2 m +M_{b}} , \nonumber \\
\mathbf{\rho} = \frac{1}{\sqrt{2}} ( \mathbf{r}_{q_{1}} -  \mathbf{r}_{q_{2}} ), \nonumber \\
\mathbf{\lambda} =\sqrt{\frac{M_{b}}{2 (2 m +M_{b})}} (
\mathbf{r}_{q_{1}} + \mathbf{r}_{q_{2}} -2 \mathbf{r}_{b } ).
\end{array}  \right.
\end{eqnarray}
In this model, the normalized wavefunction of $\Lambda_b$ has the
form
\begin{eqnarray}
\mid \Lambda_b(\mathbf{P},s) > &=& A_{B} \sum_{color,spin}
C^{s}_{s_{q_{1}},s_{q_{2}},s_{b}} \chi_{spin} \varphi_{color} \int
d^{3} \mathbf{p}_{\rho} d^{3} \mathbf{p}_{\lambda} \psi_{B_{B}}
(\mathbf{p}_{\rho},\mathbf{p}_{\lambda}) \nonumber \\ &&\mid
q_{1}(\mathbf{p}_{q_{1}},s_{q_{1}}),q_{2}(\mathbf{p}_{q_{2}},s_{q_{2}}),b(\mathbf{p}_{b},s_{b})
>,
\end{eqnarray}
thus the $\mid \Lambda_b(\mathbf{P},s) >$ satisfies the
normalization condition
\begin{equation}
<  \Lambda_b(\mathbf{P},s)   \mid \Lambda_b(\mathbf{P'},s') > = (2
\pi)^{3} \frac{M_{B_{B}} }{\omega_{P} } \delta^{3} (\mathbf{P} -
\mathbf{P'}) \delta_{s,s'}.
\end{equation}
Here $C^{s}_{s_{u},s_{d},s_{b}}$ are the Clebsch-Gordon
coefficients for combining the three spin-$\frac{1}{2}$
constituent quarks into a spin-$\frac{1}{2}$ baryon and $A_{B}$ is
a normalization factor. The spatial part of the wave function
$\psi_{B_{B}}(\mathbf{p}_{\rho},\mathbf{p}_{\lambda}) $ is the
eigenfunction of the three-body harmonic oscillator hamiltonian
and has the form
\begin{equation}
\psi_{B_{B}}(\mathbf{p}_{\rho},\mathbf{p}_{\lambda})= \exp (-
\frac{\mathbf{p}^{2}_{\rho}}{ 2 \alpha^{2}_{\rho}  } -
\frac{\mathbf{p}^{2}_{\lambda}}{ 2 \alpha^{2}_{\lambda}  }),
\end{equation}
where the parameters $\alpha_{\rho}$ and  $\alpha_{\lambda}$ are
free parameters and to be deterimined by fitting the data of
$\Lambda_b\rightarrow\Lambda_c+e^-+\bar{\nu}$.\\

(b) For the diquark picture.\\

In the diquark picture, the two light quarks constitute a
boson-like subject in the color-anti-triplet so called as diquark
and $\Lambda_b$ is supposed to be composed of the heavy quark b
and one light scalar diquark. In analog to the meson quark model,
the spatial coordinates of the two constituents in $\Lambda_b$ are
denoted by $\mathbf{r}_{b}$ and $\mathbf{r}_{D}$, and we can also
introduce the Jacobi coordinates $\mathbf{R}$ and
$\mathbf{\rho}$:
\begin{eqnarray}
\left\{
\begin{array}{c}
 \mathbf{R}=\frac{ m_{D} \mathbf{r}_{q} +
M_{b}
\mathbf{r}_{b}}{m_{D} +M_{b}} , \\
\mathbf{\rho} =\mathbf{r}_{b} -  \mathbf{r}_{D},
\end{array}  \right.
\end{eqnarray}
where $m_D$ is the diquark mass. In this model, the normalized
state vector is written as
\begin{equation}
\mid \Lambda_b^{(D)}(\mathbf{P},s) > = A_{D} \sum_{color,spin}
C^{s}_{s_{D},s_{b}} \chi_{s_{D},s_{b}} \varphi_{color} \int
d\mathbf{p}_{\rho} \psi_{B_{D}}(\mathbf{p}_{\rho}) \mid
D(\mathbf{p}_{D},s_{D}),b(\mathbf{p}_{b},s_{b}) >,
\end{equation}
where the superscript D denotes the diquark picture and $\mid
\Lambda_b^{(D)}(\mathbf{P},s)
>$ satisfies the normalization condition
\begin{equation}
< \Lambda_b^{(D)}(\mathbf{P},s)   \mid
\Lambda_b^{(D)}(\mathbf{P'},s') > = (2 \pi)^{3} \frac{M_{B_{D}}
}{\omega_{P} } \delta^{3} (\mathbf{P} - \mathbf{P'})
\delta_{s,s'}.
\end{equation}
Here $C^{s}_{s_{D},s_{b}}$ are the Clebsch-Gordon coefficients for
combining  one spin-$\frac{1}{2}$ constituent quark $b$ and one
spin-0 constituent diquark into a spin-$\frac{1}{2}$ baryon. The
corresponding wave function   $\psi_{B_{D} }(\mathbf{p}_{\rho}) $
is
\begin{equation}
\psi_{B_{D}}(\mathbf{p}_{\rho})= \exp (-
\frac{\mathbf{p}^{2}_{\rho}}{ 2 \alpha^{2}_{D}}),
\end{equation}\\
where the parameters $\alpha_{D}$ is also a free parameter and
should be determined by fitting the data of $\Lambda_b\rightarrow
\Lambda_c+e^-+\bar{\nu}$ in  the heavy-quark-light-diquark
structure.

\section{The Semi-leptonic decays of b-hadrons } \label{sec.2}
The matrix elements for the processes $B^{-} (\bar{B}^{0})
\longrightarrow D^{0} (D^{+}) e \overline{\nu}_{e}$ and
$\Lambda_{b} \longrightarrow \Lambda_{c} e \overline{\nu}_{e}$ are
expressed as
\begin{equation}
\mathcal{M}(s',s)= \frac{G_{F}}{\sqrt{2}} V_{cb} L^{\mu}
H_{\mu}(s',s)
\end{equation}\\
with the leptonic and hadronic parts being
\begin{eqnarray}
 L^{\mu}&=& \bar{u}_{e}(p_{e}) \gamma^{\mu} (1- \gamma_{5})
 \upsilon_{\nu} (p_{\nu}),\\
H_{\mu}(s,s')&=&< h_c(P,s) \mid \bar{c}(p_{c},s_{c}) \gamma_{\mu}
(1- \gamma_{5})
 b(p_{b},s_{b})  \mid h_b(P',s') >,
\end{eqnarray}
where $h_b$ and $h_c$ are corresponding b and c hadrons and $P'$,
$P$ are the four-momenta of $h_b$ and $h_c$ respectively. Defining
\begin{equation}
\widetilde{K} \equiv P'-P,
\end{equation}
we obtain
\begin{equation}
\frac{d \Gamma}{d{\bf P}} = F_{Had}   \frac{G_{F}^{2}
V_{cb}^{2}}{12 } (\widetilde{K}_{\mu} \widetilde{K}_{\nu}
-\widetilde{K}^{2} g_{\mu \nu }) \sum_{spin} T_{had}^{ \mu \nu},
\end{equation}
with
\begin{equation}
 T^{had }_{\mu \nu}=H^{\dagger}_{\mu} H_{\nu}.
\end{equation}\\

\subsection{Semi-leptonic  decays of B-mesons }
In the rest frame of the B-meson ($P'=(M_{B},\mathbf{0}$)), we
have
\begin{eqnarray}
\left\{
\begin{array}{l}
\mathbf{p}_{b} = \mathbf{p}^{b}_{\rho} , \\
\mathbf{p}^{b}_{q} = - \mathbf{p}^{b}_{\rho} ,
\end{array}  \right. \ \ \
\left\{
\begin{array}{l}
\mathbf{p_{c}}=\frac{m_{c}}{m_{q}+m_{c}} \mathbf{P} + \mathbf{p}^{c}_{\rho} , \\
\mathbf{p}^{c}_{q} = \frac{m_{q}}{m_{q}+m_{c}} \mathbf{P} -
\mathbf{p}^{c}_{\rho}.
\end{array}  \right.
\end{eqnarray}
In the semileptonic decays, the light quark is indeed a spectator,
so we have
\begin{eqnarray}
\left\{
\begin{array}{l}
\mathbf{p}^{b}_{q} = \mathbf{p}^{c}_{q} ,\\
 \mathbf{P}
=\mathbf{p_{c}}-\mathbf{p_{b}}.
\end{array}  \right.
\end{eqnarray}
and the hadronic matrix element of $B\rightarrow D+e^-+\bar{\nu}$
is
\begin{eqnarray}
 H_{\mu}&=&  A^{B}_{M} A^{D}_{M}
 \int d^{3} p^{b}_{\rho} d^{3} p^{c}_{\rho} \psi^{b*}_{M} \psi^{c}_{M}   < c_{p_{c},s_{c}},\bar{q}_{p_{q},s_{q}} \mid  \bar{c}(p_{c},s_{c})
\gamma_{\mu}
 b(p_{b},s_{b})  \mid b_{p_{b},s_{b}},\bar{q}_{p_{q},s_{q}} >,\nonumber \\
&=& (2 \pi)^{3} A^{B}_{M} A^{D}_{M}
 \int d^{3} p^{b}_{\rho} d^{3} p^{c}_{\rho} \exp{[-(\frac{p^{b }_{\rho}}{\sqrt{2} \alpha_{B}})^{2}-(\frac{p^{c }_{\rho}}{\sqrt{2} \alpha_{C}})^{2}]
 } \frac{E_{q}}{m_{q}}
 \nonumber \\&& \delta^{3} (\mathbf{p}^{c }_{\rho} -\mathbf{p}^{b}_{\rho} -\frac{m_{q}}{m_{q}+m_{c}} \mathbf{P} ) \  \bar{c}(p_{c},s_{c})
\gamma_{\mu}
 b(p_{b},s_{b}).
\end{eqnarray}
Defining
\begin{equation}
\mathbf{p}_{\rho}  \equiv \mathbf{p}^{b}_{\rho} +\frac{m_{q}
\alpha^{2}_{B}}{(m_{q}+m_{c})(\alpha^{2}_{B}+\alpha^{2}_{C})}
\mathbf{P},
\end{equation}
we obtain
\begin{eqnarray}
 H_{\mu}=  (2 \pi)^{3}  A^{B}_{M} A^{D}_{M}  \exp{[-\frac{ m^{2}_{q} P^{2}}{ 2(m_{q}+m_{c})^{2}( \alpha^{2}_{B}+ \alpha^{2}_{C})}   ]}
 \int d^{3} p_{\rho}  \frac{E_{q}}{m_{q}}  \exp{[-\frac{\alpha^{2}_{B}+\alpha^{2}_{C}}{ 2 \alpha^{2}_{B} \alpha^{2}_{C}} p^{2}_{\rho}  ]}
 \  \bar{c}_{s_{c}}
\gamma_{\mu}
 b_{s_{b}}.
\end{eqnarray}
In the non-relativistic  limit, we finally reach an expression as
\begin{equation}
\frac{d \Gamma}{d{\bf P} } = \frac{  G_{F}^{2} V_{cb}^{2}}{2
\times 192 (2 \pi)^{4} M_{B} M_{D}} (\widetilde{K}^{\mu}
\widetilde{K}^{\nu} -\widetilde{K}^{2} g^{\mu \nu }) \sum_{spin}
H^{\dagger}_{ \mu} H_{ \nu}.
\end{equation}\\

\subsection{Semi-leptonic decay of $\Lambda_b\rightarrow\Lambda_c+e^-+\bar{\nu}$ }

(a) For the 3-valence-quark model\\

In the rest frame of $\Lambda_b$ ($P'=(M_{B},\mathbf{0}$)), we
have
\begin{eqnarray}
\left\{
\begin{array}{l}
\mathbf{p_{b}}=-\sqrt{\frac{2 m_{b}}{2 m+ m_{b}}} \mathbf{p}^{b}_{\lambda} , \\
\mathbf{p}^{b}_{q_{1}} = \sqrt{\frac{ m_{b}}{2(2 m+ m_{b})}}
\mathbf{p}^{b}_{\lambda}+\frac{1}{\sqrt{2}} \mathbf{p}^{b}_{\rho} ,\\
\mathbf{p}^{b}_{q_{2}} = \sqrt{\frac{ m_{b}}{2(2 m+ m_{b})}}
\mathbf{p}^{b}_{\lambda}-\frac{1}{\sqrt{2}} \mathbf{p}^{b}_{\rho},
\end{array}  \right.
\left\{
\begin{array}{l}
\mathbf{p_{c}}=\frac{m}{2 m+ m_{c}} \mathbf{P} -\sqrt{\frac{2 m_{c}}{2 m+ m_{c}}} \mathbf{p}^{c}_{\lambda} , \\
\mathbf{p}^{c}_{q_{1}} =\frac{m}{2 m+ m_{c}} \mathbf{P}+
\sqrt{\frac{ m_{c}}{2(2 m+ m_{c})}}
\mathbf{p}^{c}_{\lambda}+\frac{1}{\sqrt{2}} \mathbf{p}^{c}_{\rho} ,\\
\mathbf{p}^{c}_{q_{2}} =\frac{m}{2 m+ m_{c}} \mathbf{P}+
\sqrt{\frac{ m_{c}}{2(2 m+ m_{c})}}
\mathbf{p}^{c}_{\lambda}-\frac{1}{\sqrt{2}} \mathbf{p}^{c}_{\rho},
\end{array}  \right.
\end{eqnarray}
and
\begin{eqnarray}
\left\{
\begin{array}{l}
\mathbf{p}^{b}_{q_{1}} = \mathbf{p}^{c}_{q_{1}} ,\\
\mathbf{p}^{b}_{q_{2}} = \mathbf{p}^{c}_{q_{2}} ,\\
 \mathbf{P}
=\mathbf{p_{c}}-\mathbf{p_{b}}.
\end{array}  \right.
\end{eqnarray}
Thus
\begin{eqnarray}
 H_{\mu}&=& (2 \pi)^{6} A_{\Lambda_{b}} A_{\Lambda_{c}} [\frac{4(2
 m+m_{c})}{m_{c}}]^{\frac{3}{2}}
  \int d^{3} p^{b}_{\lambda} d^{3} p^{b}_{\rho}  d^{3} p^{c}_{\lambda} d^{3} p^{c}_{\rho}  \frac{ E_{q_{1}} E_{q_{2}} }{m^{2}}  \nonumber \\&&
   \exp{[-(\frac{p^{b }_{\rho}}{\sqrt{2} \alpha^{b}_{\rho}})^{2}-(\frac{p^{b }_{\lambda}}{\sqrt{2} \alpha^{b}_{\lambda}})^{2}-(\frac{p^{c }_{\rho}}{\sqrt{2} \alpha^{c}_{\rho}})^{2}-(\frac{p^{c }_{\lambda}}{\sqrt{2} \alpha^{c}_{\lambda}})^{2}]
 }
 \nonumber \\&&   \delta^{3} (\mathbf{p}^{c}_{\rho} -\mathbf{p}^{b}_{\rho}) \delta^{3} (\mathbf{p}^{c}_{\lambda} -a_{1} \mathbf{p}^{b }_{\lambda} + a_{2} \mathbf{P} )\bar{c}(p_{c},s_{c})
\gamma_{\mu} (1- \gamma_{5})
 b(p_{b},s_{b}).
\end{eqnarray}
Defining
\begin{equation}
\left\{
\begin{array}{l}
\mathbf{p}_{\rho}  \equiv  \mathbf{p}^{b}_{\rho} ,\\
\mathbf{p}_{\lambda}  \equiv  \mathbf{p}^{b}_{\lambda} - \frac{m
\sqrt{2 m_{b}(2m+m_{b}) } \alpha^{b 2}_{\lambda}}{m_{c}( 2
m+m_{b})\alpha^{c 2}_{\lambda}+m_{b}( 2 m+m_{c})\alpha^{b
2}_{\lambda}} \mathbf{P},
\end{array}  \right.
\end{equation}
we have
\begin{eqnarray}
 H_{\mu}&=& (2 \pi)^{6} A_{\Lambda_{b}} A_{\Lambda_{c}} [\frac{4(2
 m+m_{c})}{m_{c}}]^{\frac{3}{2}} \exp{[-
 a_{3} P^{2}  ] }
  \int d^{3} p_{\lambda} d^{3} p_{\rho}  \frac{ E_{q_{1}} E_{q_{2}}
  }{m^{2}} \nonumber \\&&
   \exp{[-\frac{\alpha^{b2}_{\rho} + \alpha^{c2}_{\rho}}{2 \alpha^{c2}_{\rho} \alpha^{b2}_{\rho}} p^{2}_{\rho}- a_{4} p^{2}_{\lambda}]
 }
 \bar{c}(p_{c},s_{c})
\gamma_{\mu} (1- \gamma_{5})
 b(p_{b},s_{b})
\end{eqnarray}
and
\begin{eqnarray}
\left\{
\begin{array}{l}
a_{1} = [\frac{m_{b}(2 m+m_{c})}{m_{c}(2 m+m_{b})}]^{\frac{1}{2}} ,\\
a_{2} = m[\frac{2}{m_{c}(2 m+m_{c})}]^{\frac{1}{2}},\\
a_{3}=\frac{m^{2}(2m+m_{b}) }{(2m+m_{c})[m_{c}(2 m+m_{b})\alpha^{c
2}_{\lambda}+m_{b}( 2 m+m_{c})\alpha^{b 2}_{\lambda}]},\\
a_{4}=\frac{m_{c}(2 m+m_{b})\alpha^{c 2}_{\lambda}+m_{b}( 2
m+m_{c})\alpha^{b 2}_{\lambda}}{2 m_{c}(2 m+m_{b})
\alpha^{c2}_{\rho} \alpha^{b2}_{\rho}},
\end{array}  \right.
\end{eqnarray}
then the final expression is reached
\begin{equation}
\frac{d \Gamma}{d {\bf P} } = \frac{ (2 \pi)^{2} G_{F}^{2}
V_{cb}^{2}}{ 192  } (\widetilde{K}^{\mu} \widetilde{K}^{\nu}
-\widetilde{K}^{2} g^{\mu \nu }) \sum_{spin} H^{\dagger}_{ \mu}
H_{ \nu}.
\end{equation}\\

(b) For the  diquark Model\\

Now let us turn to the model for the quark-diquark structure. In
analog with the B-meson case, the semileptonic decay can be
evaluated with the two-body wavefunction and then $H_{\mu}$ is
\begin{eqnarray}
 H_{\mu}=  2 (2 \pi)^{3}  A^{b}_{D} A^{c}_{D}  \exp{[-\frac{ m^{2}_{D} P^{2}}
 { 2(m_{D}+m_{c})^{2}( \alpha^{b2}_{D}+ \alpha^{c2}_{D})}   ]}
 \int d^{3} p_{\rho}  \omega_{D}  \exp{[-\frac{\alpha^{b2}_{D}+
 \alpha^{c2}_{D}}{ 2 \alpha^{b2}_{D} \alpha^{c2}_{D}} p^{2}_{\rho}  ]}
 \  \bar{c}_{s_{c}}
\gamma_{\mu} (1- \gamma_{5})
 b_{s_{b}}
\end{eqnarray}
and
\begin{equation}
\frac{d \Gamma}{d{\bf P} } = \frac{ 4 G_{F}^{2} V_{cb}^{2}}{ 192
(2 \pi)^{4}  } (\widetilde{K}^{\mu} \widetilde{K}^{\nu}
-\widetilde{K}^{2} g^{\mu \nu }) \sum_{spin} H^{\dagger}_{ \mu}
H_{ \nu}.
\end{equation}\\

\section{Inclusive decays of b-hadrons} \label{sec.3}

(a) The spectator  contributions to the lifetimes of $\Lambda_b$,
$B^{\pm}$ and $B^0$\\

With quark-hadron duality and the optical theorem, the full
inclusive decay width {corresponding to the lifetime} of a heavy
hadron $h_b$(containing a heavy quark $b$) is related to the
absorptive part of the forward scattering matrix element
\begin{eqnarray}
\Gamma(h_{b} \rightarrow X)& =& \frac{1}{ \rho   N_{i}  }
\sum_{spin,f} \int d \prod_{f} (2 \pi)^{4} \delta^{4} (p_{i}
-p_{f} ) \mid \mathcal{M }(H_{Q} \rightarrow X ) \mid ^{2} \nonumber  \\
& =& \frac{1}{ \rho   N_{i}  }  \sum_{spin}  2 \mathcal{I}m
\mathcal{M }(H_{Q} \rightarrow H_{Q} ) \nonumber  \\
& =& \frac{2}{\rho N_{i}}Im\int\!\!d^{4}x<H_{Q} \mid \hat{T}\mid
H_{Q}>  = \frac{ 2}{ \rho N_{i} }     < H_{Q} \mid \hat{\Gamma}
\mid H_{Q} >
\end{eqnarray}
where
\begin{equation}
\hat{T}=T\{i \mathcal{L}_{eff}(x),\mathcal{L}_{eff}(0)\},
\end{equation}
and $\rho  , N_{i} $ are the state density factors and
$\mathcal{L}_{eff}$ is the relevant effective weak Lagrangian
which is responsible for the decay. For the concerned final state
X with the designated quark-antiquark combination, up-to order
$1/m^{3}_{Q}$ one finds  \cite{Bigi,N.Bilic}:
\begin{eqnarray}
\Gamma(H_{Q} \rightarrow X)&=&\frac{G^{2}_{F} m^{2}_{Q}}{192
\pi^{3} } \mid V(CKM) \mid^{2} \{ c_{3}^{X} <H_{Q} \mid  \bar{Q} Q
\mid H_{Q}> +  c_{5}^{X} \frac{<H_{Q} \mid  \bar{Q} i \sigma \cdot
G Q  \mid H_{Q}>}{m^{2}_{Q}}  \nonumber \\ & + & \sum_{i}
c_{6,i}^{X}
 \frac{<H_{Q} \mid  ( \bar{Q} \Gamma_{i} q ) ( \bar{q} \Gamma_{i} Q )
 \mid H_{Q}>}{m^{3}_{Q}} +  \mathcal{O}(1/m^{4}_{Q}) \}.
\end{eqnarray}
Here only the heavy quark (b,c quark) decays are concerned. In the
spectator scenario, the light flavor(s) in the heavy meson
(baryon) remains as a spectator.  We can write the spectator
contribution as
\begin{equation}
\Gamma_{spec}=\sum_{l=e,\mu,\tau} \Gamma_{b \rightarrow c l
\bar{\nu}_{l} } + \sum_{q=u,d,s,c} \Gamma_{b \rightarrow c \bar{q}
q}.
\end{equation}
The pure b-quark decay rates of $b \rightarrow c \bar{u} s $, $ c
\bar{u} d +c e \bar{\nu}_{e}$,$c \mu \bar{\nu}_{\mu}$  and $c \tau
\bar{\nu}_{\tau}$ have been carefully evaluated by Bagan et al.
\cite{Bagan} as
\begin{eqnarray}
\Gamma(b \rightarrow c \bar{u} s  + c \bar{u} d ) &=& (4.0 \pm 0.4
)
\Gamma_{b \rightarrow c e \bar{\nu}_{e}} , \nonumber \\
\Gamma(b \rightarrow c \tau \bar{\nu}_{\tau} ) &=& 0.25 \Gamma_{b
\rightarrow c e \bar{\nu}_{e}} ,
\end{eqnarray}
and the measured  semileptonic branching ratio is given in
ref.\cite{Bagan} as $B(b \rightarrow c e \bar{\nu}_{e}) = (11.6
\pm 1.8) \%  $. The theoretically formula at the tree level reads
\begin{equation}
\Gamma(b \rightarrow c e \bar{ \nu}_{e})=\mid V_{cb} \mid^{2}
\frac{G_{F}^{2} m^{5}_{b}}{192 \pi^{3}} [1 - 8
(\frac{m_{c}}{m_{b}})^{2} -12 (\frac{m_{c}}{m_{b}})^{4} \ln
\frac{m^{2}_{c}}{m^{2}_{b}} + 8 (\frac{m_{c}}{m_{b}})^{6} -
(\frac{m_{c}}{m_{b}})^{8} ].
\end{equation}\\
 The semi-leptonic and non-leptonic decay rates of b-quark up-to
order $1/m_{b}^{2}$ have been evaluated by many authors
\cite{Bigi}. In our numerical computations we need to use their
formulas, for the readers' convenience we collect them in Appendix
A.\\

(b) The non-spectator  contributions to the inclusive decays of
b-hadrons.\\

The total width of a b-hadron decay (the lifetime) includes two
contributions
\begin{equation}
\Gamma(H_{Q } \rightarrow X )=\Gamma_{b}^{spec}+\Gamma_{b}^{non},
\end{equation}
where the superscripts $spec$ and $non$ refer to the spectator and
non-spectator contributions respectively.
 To estimate the non-spectator contribution to  b-hadron
decays, the basis is the weak effective Lagrangian \cite{Neubert},
\begin{eqnarray}
\mathcal{H}_{eff}=\frac{4G_{F}}{\sqrt{2}}  V_{cb} &\{&
c_{1}(m_{b})[ \bar{d'} \gamma^{\mu} L  u  \bar{c} \gamma_{\mu} L b
+ \bar{s'}
\gamma^{\mu} L  c   \bar{c} \gamma_{\mu} L  b] \nonumber \\
 &+& c_{2}(m_{b})[
\bar{c} \gamma^{\mu} L  u  \bar{d'} \gamma_{\mu} L  b + \bar{c}
\gamma^{\mu} L  c   \bar{s'} \gamma_{\mu} L  b]   \nonumber \\
&+& \sum_{l=e,\mu,\tau} \bar{l} \gamma^{\mu} L  \nu_{e}  \bar{c}
\gamma_{\mu} L  b \},
\end{eqnarray}
where $d'=d \cos \theta _{c} + s \sin \theta _{c} $,  $s'=s \cos
\theta _{c} -d \sin \theta _{c} $ and $\theta_C$ is the Cabibbo
angle. $\mathcal{H}_{eff}$ can be further written as
\begin{eqnarray}
\mathcal{H}_{eff}=\frac{4G_{F}}{\sqrt{2}} & [ & V^{\ast}_{ub}
V_{uq} ( c_{1u}(\mu) O^{u}_{1} + c_{2u}(\mu) O^{u}_{2})  +
V^{\ast}_{cb} V_{uq} ( c_{1c}(\mu) O^{c}_{1} + c_{2c}(\mu)
O^{c}_{2})\nonumber \\& -& V^{\ast}_{tb} V_{tq} \sum^{6}_{i=3}
c_{i}(\mu) O_{i} ],
\end{eqnarray}
where the operations are
\begin{eqnarray}
O^{Q}_{1} &=& \bar{Q} \gamma^{\mu} L  b \bar{q} \gamma_{\mu}L  u   ,  \nonumber        \\
O^{Q}_{2} &=& \bar{Q}_{i} \gamma^{\mu} L  b_{j}
\bar{q}_{j} \gamma_{\mu} L  u_{i}   , \nonumber  \\
O_{3,5} &=& \bar{q} \gamma^{\mu} L  b \sum_{q'} \bar{q'} \gamma_{\mu} L (R) q'   , \nonumber        \\
O_{4,6} &=& \bar{q}_{i} \gamma^{\mu} L  b_{j} \sum_{q'}
\bar{q'}_{j} \gamma_{\mu} L (R) q'_{i}    ,
\end{eqnarray} \\
and $L(R)=\frac{1 \mp \gamma_{5}}{2}$, $q'$ is summed over
$u,d,s,c,b$ and $t$,  $q$ can be $s $ or $d$, $i$
and $ j$ are the color indices.\\

(c) Inclusive decays of B-mesons\\

The non-spectator operators for  B-meson decays have been given
\cite{Neubert}, because of the Cabibbo suppression  the W-exchange
(WE) process only exists for $B^{0}$ and the PI-process applies
uniquely to $B^{\pm}$. One can have \cite{Bigi}
\begin{eqnarray}
\hat{\Gamma}^{B^{\pm}}_{PI}& =&  \frac{2 G^{2}_{F} m^{2}_{b} }{
\pi } \mid V_{cb}  \mid^{2}(1- z)^{2} \{[2 c_{1}c_{2}+ \frac{1
}{N_{c} }(c^{2}_{1}+c^{2}_{2}) ] O^{u}_{V-A} + 2
(c^{2}_{1}+c^{2}_{2})T^{u}_{V-A}) \},       \\
\hat{\Gamma}^{B^{0}}_{WE}& =& - \frac{2 G^{2}_{F} m^{2}_{b} }{ 3
\pi } \mid V_{cb}  \mid^{2}(1- z)^{2} \{(2 c_{1}c_{2}+ \frac{1
}{N_{c} } c^{2}_{1}+ N_{c} c^{2}_{2} )\nonumber  \\
&& [(1+\frac{z}{2}) O^{d'}_{V-A} -(1+2z) O^{d'}_{S-P} ] + 2
c^{2}_{1}[(1-z)T^{s'}_{V-A} -(1+2z) T^{s'}_{S-P}] \},
\end{eqnarray} \\
where $z=\frac{m^{2}_{c}}{m^{2}_{b}}$ and $N_{c}=3$. The local
four-quark operators appearing in this expression are
\begin{eqnarray}
O^{q}_{V-A} &=& \bar{b} \gamma^{\mu} L  q \bar{q} \gamma_{\mu}L  b   ,  \nonumber        \\
O^{q}_{S-P} &=& \bar{b}  L  q
\bar{q} R  b   , \nonumber  \\
T^{q}_{V-A} &=& \bar{b} \gamma^{\mu} L T^{a} q \bar{q} \gamma_{\mu} L T^{a} b   ,  \nonumber        \\
T^{q}_{S-P} &=& \bar{b}  L T^{a} q \bar{q} R T^{a}  b   ,
\end{eqnarray} \\
where $T^{a}=\frac{\lambda_{a}}{2}$ are the generators of the
color SU(3). The hadronic matrix element is
\begin{eqnarray}
\Gamma^{B_{q}}_{non} &=&\frac{1}{M_{B_{q}}} < B_{q}
\mid \hat{\Gamma}^{B_{q}}_{non} \mid B_{q} > \nonumber \\
& =& \frac{1}{M_{B_{q}}} \mid A_{M} \mid^{2}  \sum_{color , spin}
\int d^{3} \mathbf{p'}_{\rho}  d^{3} \mathbf{p}_{\rho}
\psi^{\ast}_{B_{q}}(\mathbf{p'}_{\rho})
\psi_{B_{q}}(\mathbf{p}_{\rho}) \nonumber
\\ && < \bar{q}(\mathbf{p'}_{q},s'_{q}),b(\mathbf{p'}_{b},s'_{b}) \mid
\hat{\Gamma}^{B_{q}}_{non} \mid
q(\mathbf{p}_{q},s_{q}),b(\mathbf{p}_{b},s_{b})
>.
\end{eqnarray}
It is reasonable to assume that the quarks in the bound states are
only slightly off-shell, we can carry out the calculation of the
matrix elements in our model described in last section.
\begin{eqnarray}
\Gamma^{B^{\pm}}_{PI} &=&   \frac{3 G^{2}_{F} m^{2}_{b} }{ \pi
M_{B^{0}} } \mid V_{cb}  \mid^{2} [2 c_{1}c_{2}+ \frac{1 }{N_{c}
}(c^{2}_{1}+c^{2}_{2}) ](1- z)^{2} \mid A_{B^{\pm}} \mid^{2}
\nonumber
\\ &&  \sum_{
spin} \int d^{3} \mathbf{p'}_{\rho} d^{3} \mathbf{p}_{\rho}  \exp
[-\frac{p'^{2}_{\rho} +p^{2}_{\rho}}{ 2 \alpha^{2}_{B^{\pm}}}] \
 \bar{b} \gamma^{\mu} L u \bar{u} \gamma_{\mu}L b,
\end{eqnarray}
\begin{eqnarray}
\Gamma^{B^{0}}_{WE} &=&  - \frac{ G^{2}_{F} m^{2}_{b} }{ \pi
M_{B^{0}} }  \mid  V_{cb} \cos{\theta_{c}} \mid^{2} [2 c_{1}c_{2}+
\frac{c^{2}_{1} }{N_{c} }+ N_{c} c^{2}_{2} ](1- z)^{2} \mid
A_{B^{0}} \mid^{2} \sum_{ spin} \int d^{3} \mathbf{p'}_{\rho}
d^{3} \mathbf{p}_{\rho} \nonumber
\\ &&  \exp
[-\frac{p'^{2}_{\rho} +p^{2}_{\rho}}{ 2 \alpha^{2}_{B^{0}}}] \{
(1+\frac{z}{2}) \bar{b} \gamma^{\mu} L d \bar{d} \gamma_{\mu} L b
+(1+ 2 z) \bar{b}  L  d \bar{d} L  b \}.
\end{eqnarray}\\

(d) The inclusive  Decays of $\Lambda_b$\\

(1)  For the 3-valence-quark model\\

The concerned non-spectator Feynman diagrams are shown in Fig.1
(a), (b) and (c). One of the light valence quarks and the b-quark
exchange a W-boson while the other light quark stands by as a
spectator. Fig.1 (c) is a PI diagram\footnote{As discussed in the
previous footnote, in fact, this PI diagram is not exactly the
Pauli Interference diagrams discussed in the literature. In the
diagram a quark from the initial state crosses with one from the
final state. One can see that a quark from the "right" state joins
the "left" vertex and vice versa, thus for such a situation, we
just keep the terminology as "PI" diagrams.} where three quarks
take part in the reaction. Due to the CKM entries, other diagrams
are much suppressed compared to that in Fig1.(a),(b) and (c), thus
can be neglected.

\begin{eqnarray}
\hat{\Gamma}^{\Lambda_{b}}_{WE}& =&  \frac{G^{2}_{F} M^{2}_{+}
}{\pi } \mid V_{cb} V_{uq}  \mid^{2}(1- z_{+})(1-8 z_{+} ) \{
[c^{2}_{1c}+c^{2}_{2c}+ \frac{2 c_{1c}c_{2c} }{N_{c} }]
O^{c}_{V-A} + 4 c_{1c}c_{2c} T^{c}_{V-A}) \},\nonumber
\\
\hat{\Gamma}^{\Lambda_{b}}_{PI}& =&   G^{2}_{F}  \mid V_{cb}V_{ud}
\mid^{2}  \frac{4 m_{c} \Gamma_{c}  }{ (p^{2}_{c} -m^{2}_{c} )^{2}
+  m^{2}_{c} \Gamma^{2}_{c} } \{ c^{2}_{1c }
\widetilde{O}_{V-A}(b_{i},b_{i},u_{j},d_{j},d_{k},u_{k}) \nonumber
\\ & & +c^{2}_{2c }
\widetilde{O}_{V-A}(b_{i},b_{j},u_{k},d_{i},d_{j},u_{k}) +2 c_{1c}
c_{2c} \widetilde{O}_{V-A}(b_{i},b_{j},u_{k},d_{k},d_{j},u_{i})
 \}  \nonumber
\\ & &  ( g^{\mu \lambda} p^{\nu}_{c} - g^{\nu \lambda} p^{\mu}_{c}
+ g^{\mu \nu } p^{\lambda}_{c} - i \varepsilon^{\mu \nu \lambda
\rho} p_{ c \rho} ) ,
\end{eqnarray}
where $z_{+}=\frac{m^{2}_{c}}{M^{2}_{+}}$ and $M^{2}_{+}=(p_{b} +
p_{u})^{2}
 \approx m^{2}_{b} $, $p_{c}=p_{b}+p_{u}-p_{d}$
and the relevant four-quark operators are
\begin{eqnarray}
 O^{c}_{V-A}& = & \bar{b} \gamma^{\mu} L b \bar{u} \gamma_{\mu} L u  , \nonumber \\
 T^{c}_{V-A}& = & \bar{b} \gamma^{\mu} L T^{a} b \bar{u} \gamma_{\mu} L T^{a} u  , \nonumber \\
 \widetilde{O}_{V-A} & = & \bar{b} \gamma_{\mu} L b \bar{u} \gamma_{\nu} L d \bar{d} \gamma_{\lambda} L
 u.
 \end{eqnarray}
The hadronic matrix elements are evaluated in the non-relativistic
harmonic oscillator model as
\begin{eqnarray}
 \Gamma_{WE}&=&  < \Lambda_{b} (\mathbf{P}=\mathbf{0}) \mid \hat{\Gamma}_{WE}   \mid \Lambda_{b}
(\mathbf{P'}=\mathbf{0}) > \nonumber \\
& = & (2 \pi)^{3} \mid A_{\Lambda_{b}} \mid^{2} \sum_{color ,
spin} \int d^{3} \mathbf{p'}_{\rho} d^{3}\mathbf{p'}_{\lambda}
d^{3} \mathbf{p}_{\rho} d^{3} \mathbf{p}_{\lambda} \frac{E_{d}}{m}
\delta^{3}( \mathbf{p'}_{d}-\mathbf{p}_{d} )   \nonumber \\
& & \exp [-\frac{p'^{2}_{\rho}+p^{2}_{\rho}}{ 2
\alpha^{2}_{\rho}}-\frac{p'^{2}_{\lambda}+p^{2}_{\lambda}}{ 2
\alpha^{2}_{\lambda}} ] <
u(\mathbf{p'}_{u},s'_{u}),b(\mathbf{p'}_{b},s'_{b}) \mid
\hat{\Gamma }_{WE} \nonumber \\
&& \mid (\mathbf{p}_{u},s_{u}),b(\mathbf{p}_{b},s_{b})
>.
\end{eqnarray}
Setting
\begin{eqnarray}
\mathbf{k}  = \mathbf{p}_{b} - \mathbf{p}'_{b},
\end{eqnarray}
we have
\begin{eqnarray}
\left\{
\begin{array}{l}
\mathbf{p'}_{\rho}  = \mathbf{p}_{\rho} +\frac{1}{\sqrt{2}} \mathbf{k} ,\nonumber \\
\mathbf{p'}_{\lambda}  = \mathbf{p}_{\lambda} +\sqrt{\frac{2 m +
m_{b} }{2m_{b}}} \mathbf{k} ,  \end{array} \right.
\end{eqnarray}
thus
\begin{eqnarray}
\left\{
\begin{array}{l}
\widetilde{\mathbf{p}}_{\rho}  = \sqrt{2} \mathbf{p}_{\rho} +\frac{1}{2} \mathbf{k} ,\nonumber \\
\widetilde{\mathbf{p}}_{\lambda}  =\sqrt{2} \mathbf{p}_{\lambda} +
\frac{1}{2} \sqrt{\frac{2 m + m_{b} }{m_{b}}} \mathbf{k}.
\end{array} \right.
\end{eqnarray}
We finally achieve
\begin{eqnarray}
 \Gamma_{WE} &=&  (2 \pi)^{2} \frac{ G^{2}_{F} M^{2}_{+}}{4} \mid V_{cb} V_{uq}  \mid^{2} (c_{1c}-c_{2c})^{2} (1- z_{+})(1-8 z_{+} ) [\frac{9 (2m +m_{b})}{8 m_{b}}]^{\frac{3}{2}}  \mid A_{\Lambda_{b}} \mid^{2} \nonumber \\
& &   \  \sum_{ spin} \int d^{3} \widetilde{\mathbf{p}}_{\rho}
d^{3} \widetilde{\mathbf{p} }_{\lambda} d^{3} k \frac{E_{d}}{m}
\exp [-\frac{\widetilde{p}^{2}_{\rho}}{ 2
\alpha^{2}_{\rho}}-\frac{\widetilde{p}^{2}_{\lambda}}{ 2
\alpha^{2}_{\lambda}} -\frac{m_{b} \alpha^{2}_{\lambda}+ (2m
+m_{b})\alpha^{2}_{\rho}}{ 8 m_{b} \alpha^{2}_{\rho}
\alpha^{2}_{\lambda}}  k^{2} ] \nonumber \\
& &
  \bar{b} \gamma^{\mu} (1- \gamma_{5})  b \bar{u}
\gamma_{\mu} (1- \gamma_{5}) u
 ,
\end{eqnarray}
and
\begin{eqnarray}
 \Gamma_{PI} &=&  < \Lambda_{b} (\mathbf{P}=\mathbf{0}) \mid \hat{\Gamma}_{PI}   \mid \Lambda_{b}
(\mathbf{P'}=\mathbf{0}) > \nonumber \\
& = & \frac{ G^{2}_{F} }{4} \mid V_{cb} V_{ud} \mid^{2}
(c_{1c}-c_{2c})^{2} \mid A_{\Lambda_{b}} \mid^{2} \sum_{ spin}
\int d^{3} \mathbf{p'}_{\rho} d^{3}\mathbf{p'}_{\lambda}  d^{3}
\mathbf{p}_{\rho}
d^{3} \mathbf{p}_{\lambda}  \nonumber \\
& & \frac{m_{c} \Gamma_{c} }{ (p^{2}_{c} -m^{2}_{c} )^{2} +
m^{2}_{c} \Gamma^{2}_{c} } \exp
[-\frac{p'^{2}_{\rho}+p^{2}_{\rho}}{ 2
\alpha^{2}_{\rho}}-\frac{p'^{2}_{\lambda}+p^{2}_{\lambda}}{ 2
\alpha^{2}_{\lambda}} ]      \nonumber
\\ & &  \bar{b} \gamma_{\mu} (1- \gamma_{5})  b \bar{u} \gamma_{\nu} (1- \gamma_{5})  d \bar{d} \gamma_{\lambda}(1- \gamma_{5})  u   ( g^{\mu \lambda} p^{\nu}_{c} - g^{\nu \lambda} p^{\mu}_{c}
+ g^{\mu \nu } p^{\lambda}_{c} - i \varepsilon^{\mu \nu \lambda
\rho} p_{ c \rho} ).
\end{eqnarray}\\

(2) The quark-diquark structure\\

(i) Now we consider the non-spectator effects in the quark-diquark
scenario. In this picture, b-quark and the light diquark exchange
a W-boson as
\begin{equation}
b + D \rightarrow c +D',
\end{equation}
where $D$ and $D'$ are scalar or vector diquarks. Supposing that
the diquarks of $\Lambda_b$ and $\Lambda_c$ are in ground states,
i.e. $D$ and $D'$ are scalars only. The corresponding Feynman
diagrams for  the non-spectator contributions to the inclusive
decay rate of $\Lambda_b$ are shown in Fig.2. The effective
diquark-W-boson interaction vertex was given\cite{Dosch}
\begin{equation}
V_{s} = - i G_{s} ( q_{1} + q_{2})^{\mu}W_{\mu}  , \ \ \ \ \ for \
\ \ DWD'
\end{equation}\\
where $D$ and $D'$ stand for scalar diquarks, $q_{1}$ and $q_{2}$
are the four-momenta of $D$,$D'$, $G_{s}$ is a form factor which
is determined by fitting data \cite{Dosch} and generally one can
recast it as
\begin{equation}
G_{s} = \frac{ g_{s}}{2 \sqrt{2}} F_{s}(Q^{2}),
\end{equation}
with
\begin{eqnarray}
F_{s}(Q^{2}) = \frac{ \bar{\alpha}_{s}(Q^{2}) Q^{2}_{0}
}{Q^{2}_{0} + Q^{2}}.
\end{eqnarray}
The form factor represents the inner structure of the diquark.
$\bar{\alpha}_{s}(Q^{2})$ is the effective non-perturbative QCD
coupling constant and takes a reasonable value similar to that
used in the potential model. The transition amplitude for
$b(p_{1}) + D(q_{1}) \rightarrow c(p_{2}) +D'(q_{2})$ via the
W-boson exchange reads as
\begin{equation}
T^{s}_{eff} = \frac{ G_{F}}{\sqrt{2}} (V_{cb} V^{\ast}_{ud})
\bar{c} \gamma_{\mu} (1- \gamma_{5}) b (q_{1}+q_{2})^{\mu}
F_{s}(Q^{2}).
\end{equation}

The same process can occur via the penguin-induced effective
vertex. The transition $b(p_{1}) + D(q_{1}) \rightarrow c(p_{2})
+D'(q_{2})$ via the penguin, where a virtual gluon is exchanged
between the quark and diquark,  and the formulation is similar to
that for W-boson exchange. The effective vertex for $b \rightarrow
s + g$ is given  \cite{Hou} as
\begin{equation}
V^{a}_{\mu} = \frac{ G_{F}}{\sqrt{2}}  \frac{ g_{s}}{4
\pi^{2}}(V_{tb} V^{\ast}_{ts}) \bar{s}t^{a}[ \triangle F_{1}
(q^{2} \gamma_{\mu} - q_{\mu} q \!\!\!/ ) L - F_{2} i \sigma_{\mu
\nu } q^{\nu} m_{b} R] b ,
\end{equation} \\
with $\triangle F_{1}=F^{t}_{1} - F^{c}_{1},\;\;F^{t}_{1} \approx
0.25,\;\;F^{c}_{1} = -\frac{2}{3} \ln \frac{m^{2}_{c}}{M^{2}_{W}}
\approx 5.3, F_{2} \approx 0.2 $.\\
Thus we can ignore the $F_{2}$ part and  the transition amplitude
is
\begin{equation}
T^{s}_{eff} = \frac{ G_{F}}{\sqrt{2}} \frac{ \alpha_{s}}{\pi}
(V_{tb} V^{\ast}_{ts}) \bar{s}_{i} t^{a}_{ij} \gamma_{\mu} b_{j}
t^{a}_{l m } \triangle F_{1}  F_{s}(Q^{2}) (q_{1}+q_{2})^{\mu} .
\end{equation}

(ii) The  effective operators for the inclusive processes of
$\Lambda_{b}$\\

The weak  W-exchange (WE) operator is
\begin{eqnarray}
\hat{\Gamma}_{WA}^{tree} = \frac{ G^{2}_{F} \widetilde{M}_{+}^{2}
}{ 8 \pi} \mid V_{cb} V_{ud} \mid^{2} (1-\widetilde{z}_{+})
 \bar{b} [(\frac{\widetilde{M}_{+}^{2}}{2} -m^{2}_{c})(p_{D}\!\!\!\!\!\!\!/ \ +p'_{D} \!\!\!\!\!\!\!/ \ )
 +(\frac{3 \widetilde{M}_{+}^{2}}{4} -3m^{2}_{c})p_{+}\!\!\!\!\!\!\!/ \
 +\frac{1}{2} p'_{D}\!\!\!\!\!\!\!/ \ \   p_{+}\!\!\!\!\!\!\!/ \  \ p_{D}\!\!\!\!\!\!\!/
 \ \ ]Lb F^{2}_{s},
 \end{eqnarray}
where $p_{b}$, $p'_{b}$, $p_{D}$ and $p'_{D}$  are the four
momenta of the initial and final b-quarks and diquarks, $p_{+}=
p_{b} + p_{D} $, $\widetilde{M}_{+}^{2}= (p_{b} + p_{D})^{2}
\approx
 M^{2}_{\Lambda_{b}}$ and
 $\widetilde{z}_{+}=\frac{m^{2}_{c}}{\widetilde{M}_{+}^{2}}$.

For the Pauli-interference operators  (PI), it is noted that as we
indicated in the previous footnote that Pauli interference here is
only a terminology for such classes of Feynman diagrams, because
the duquark is a boson-like subject, so by no means needs to obey
the Pauli principle.

There are contributions from both the tree and penguin mechanisms
to the Pauli interference (PI) operators. In non-relativistic
limit, $Q^{2}_{i} \sim Q^{2}_{f} \sim 0$,and $F_{s}  \sim
\bar{\alpha}_{s}(Q^{2})$, they are
 \begin{eqnarray}
 \hat{\Gamma}_{PI}^{tree} = \frac{ G^{2}_{F} \widetilde{M}_{-}^{2} }{ 8
\pi} \mid V_{cb} V_{ud} \mid^{2} (1-\widetilde{z}_{-})
 \bar{b} [ ( \frac{3 \widetilde{M}_{-}^{2}}{4} -3m^{2}_{c}) p_{-}\!\!\!\!\!\!\!/
  \
  - ( \frac{\widetilde{M}_{-}^{2}}{2} -m^{2}_{c})(p_{D}\!\!\!\!\!\!\!/ \ +p'_{D} \!\!\!\!\!\!\!/ \
 )
 + \frac{1}{2} p_{D}\!\!\!\!\!\!\!/ \ \   p_{-}\!\!\!\!\!\!\!/ \  \ p'_{D}\!\!\!\!\!\!\!/
 \ \ ]Lb F^{2}_{s},
 \end{eqnarray}
 \begin{eqnarray}
\hat{\Gamma}_{PI}^{penguin} =\frac{  G^{2}_{F} }{(4 \pi)^{3}}
\alpha^{2}_{s} \mid V_{tb} V_{ts} \mid^{2}  \bar{b}_{k} [
\widetilde{M}_{-}^{2} (\frac{3}{2 } p_{-}\!\!\!\!\!\!\!/ \ \ -
p'_{D}\!\!\!\!\!\!\!\!/ \ \ - p_{D}\!\!\!\!\!\!\!/ \
 ) + p_{D}\!\!\!\!\!\!\!\!/ \ \ \  p_{-}\!\!\!\!\!\!\!/ \  \ p'_{D}\!\!\!\!\!\!\!/
 \ \ ] b_{j} t^{a}_{ki} t^{a}_{mn} t^{b}_{ij} t^{b}_{nl}
(\triangle F_{1})^{2} F^{2}_{s} ,
\end{eqnarray}
where $p_{-} = p_{b} - p'_{D} $. $\widetilde{M}_{-}^{2}= (p_{b} -
p_{D})^{2} \simeq
 (m_{b}-m_{D})^{2}$ and
 $\widetilde{z}_{-}=\frac{m^{2}_{c}}{\widetilde{M}_{-}^{2}}$.
In the operators, we set the current  quark mass of s-quark to be
zero.  Sandwiching the operators between initial and final
$\Lambda_b$ states of the forward scattering, we obtain the
hadronic matrix elements as
\begin{eqnarray}
\Gamma_{WE}^{tree} & =&   < \Lambda_{b}(\mathbf{P}=\mathbf{0},s)
\mid \hat{\Gamma}_{WE}^{tree} \mid
\Lambda_{b}(\mathbf{P}=\mathbf{0},s ) > \nonumber \\ & =& \frac{
G^{2}_{F} \widetilde{M}_{+}^{2} }{ 8 \pi} \mid V_{cb} V_{ud}
\mid^{2} (1-\widetilde{z}_{+}) \mid A_{\Lambda_{b}} \mid^{2}
\sum_{spin} \int d^{3} \mathbf{p'}_{\rho} d^{3}\mathbf{p}_{\rho}
\exp [-\frac{p'^{2}_{\rho} +p^{2}_{\rho}}{ 2 \alpha^{2}_{D}}]
 \nonumber \\ &&
 \bar{b} [(\frac{\widetilde{M}_{+}^{2}}{2} -m^{2}_{c})(p_{D}\!\!\!\!\!\!\!/ \ +p'_{D} \!\!\!\!\!\!\!/ \ )
 +(\frac{3 \widetilde{M}_{+}^{2}}{4} -3m^{2}_{c})p_{+}\!\!\!\!\!\!\!/ \
 +\frac{1}{2} p'_{D}\!\!\!\!\!\!\!/ \ \   p_{+}\!\!\!\!\!\!\!/ \  \ p_{D}\!\!\!\!\!\!\!/
 \ \ ]Lb F^{2}_{s},
 \end{eqnarray}
 \begin{eqnarray}
\Gamma_{PI}^{tree} & =&   < \Lambda_{b}(\mathbf{P}=\mathbf{0},s)
\mid \hat{\Gamma}_{PI}^{tree} \mid
\Lambda_{b}(\mathbf{P}=\mathbf{0},s ) > \nonumber
\\ & =& \frac{ G^{2}_{F} \widetilde{M}_{-}^{2} }{ 8 \pi} \mid V_{cb}
V_{ud} \mid^{2} (1-\widetilde{z}_{-}) \mid A_{\Lambda_{b}}
\mid^{2} \sum_{spin} \int d^{3} \mathbf{p'}_{\rho}
d^{3}\mathbf{p}_{\rho} \exp [-\frac{p'^{2}_{\rho} +p^{2}_{\rho}}{
2 \alpha^{2}_{D}}]
 \nonumber \\ &&
 \bar{b} [ (\frac{3 \widetilde{M}_{-}^{2}}{4} -3m^{2}_{c})p_{-}\!\!\!\!\!\!\!/ \
 -(\frac{\widetilde{M}_{-}^{2}}{2} -m^{2}_{c})(p_{D}\!\!\!\!\!\!\!/ \ +p'_{D} \!\!\!\!\!\!\!/ \ )
 +\frac{1}{2} p_{D}\!\!\!\!\!\!\!/ \ \   p_{-}\!\!\!\!\!\!\!/ \  \ p'_{D}\!\!\!\!\!\!\!/
 \ \ ]Lb F^{2}_{s},
 \end{eqnarray}
\begin{eqnarray}
\Gamma_{PI}^{penguin} & =&   <
\Lambda_{b}(\mathbf{P}=\mathbf{0},s) \mid
\hat{\Gamma}_{PI}^{penguin} \mid
\Lambda_{b}(\mathbf{P}=\mathbf{0},s ) > \nonumber \\ &= & \frac{
11 G^{2}_{F} }{18 (4 \pi)^{3}} \alpha^{2}_{s} \mid V_{tb} V_{ts}
\mid^{2} \mid A_{\Lambda_{b}} \mid^{2} \sum_{spin} \int d^{3}
\mathbf{p'}_{\rho} d^{3}\mathbf{p}_{\rho} \exp
[-\frac{p'^{2}_{\rho} +p^{2}_{\rho}}{ 2 \alpha^{2}_{D}}]
 \nonumber \\ && \bar{b} [
\widetilde{M}_{-}^{2} (p'_{D}\!\!\!\!\!\!\!/ \ \ +
p_{D}\!\!\!\!\!\!/ \ + \frac{3}{2 } p_{-}\!\!\!\!\!\!\!/ \ \
 ) + p_{D}\!\!\!\!\!\!\!/ \ \   p_{-}\!\!\!\!\!\!\!/ \  \ p'_{D}\!\!\!\!\!\!\!/
 \ \ ] b
(\triangle F_{1})^{2} F^{2}_{s}.
\end{eqnarray} \\

\section{The Numerical Results}\label{sec.4}
To evaluate the lifetimes of $B^{0}$, $B^{\pm}$ and $\Lambda_{b}$,
the input parameters  are taken as follows
\begin{eqnarray}
 V_{ud}&=&0.9742, V_{us}=0.219, V_{tb}=0.9993,
V_{ts}=0.035, V_{cb}=0.037,\nonumber \\ N_{c}&=&3.0, m_{b}=4.79
{\rm GeV}, m_{c}=1.25 {\rm GeV}, m_{u}=m_{q}=0.40 {\rm GeV},
\nonumber
\\
m_{D}&=&0.6 {\rm GeV},  m_{B^{0}}=5.28 {\rm GeV}, m_{B^{\pm}}=5.28
{\rm GeV}, m_{\Lambda_{b}}=5.6 {\rm GeV}, \nonumber
\end{eqnarray}
and the current quark masses
$\bar{m}_{u}=\bar{m}_{d}=\bar{m}_{s}=0$. In addition, in the
diquark model, we take the values given in ref.\cite{Dosch},
\begin{eqnarray}
\triangle F_{1} = -5.05;\;\; Q^{2}_{0} = 3.22 {\rm GeV}^{2};\;\;
\alpha_{s}(m^{2}_{b})=0.246;\;\;\overline{\alpha}_{s}=0.87.
\nonumber
\end{eqnarray}
Meanwhile the pure b-quark decay rate has been  evaluated by Bagan
et al \cite{Bagan} as
\begin{equation}
\Gamma_{b}=4.13 \times 10^{-13} {\rm GeV}.
\end{equation}\\

(a) For the heavy B-mesons\\

The experimental data of $B^{0}$ and $B^{-}$ are \cite{Data}
\begin{eqnarray}
\left\{
\begin{array}{rl}
\tau_{B^{0}}\!\!\!\!&=(1.548 \pm 0.032) ps, \\
 BR(\bar{B}^{0}
\rightarrow D^{+} e \bar{\nu}_{e})\!\!\!\!&= (2.10 \pm 0.19)\%  ,\\
\Gamma_{SL}(\bar{B}^{0} \rightarrow D^{+} e
\bar{\nu}_{e})\!\!\!\!&= (8.121 \sim 9.737) \times 10^{-15} {\rm
GeV}.
\end{array}
\right.
\end{eqnarray}
and
\begin{eqnarray}
\left\{
\begin{array}{rl}
\tau_{B^{\pm}}\!\!\!\!&=(1.653 \pm 0.028) ps,\\
 BR(B^{-} \rightarrow D^{0} e \bar{\nu}_{e})\!\!\!\!&= (2.15 \pm 0.22)\%  ,\\
\Gamma_{SL}(B^{-} \rightarrow D^{0} e \bar{\nu}_{e})\!\!\!\!&=
(7.685 \sim 9.347) \times 10^{-15} {\rm GeV}.
\end{array}
\right.
\end{eqnarray}

Fitting the data of the semi-leptonic  decays of B, we obtain the
corresponding  parameters  $\alpha_{B^{0}}$ and $\alpha_{B^{\pm}}$
in the wavefunctions as
\begin{eqnarray}
 \alpha_{B^{0}}=0.573    ;   \alpha_{B^{\pm}}=0.541   ; \nonumber \\
\end{eqnarray}
With the Wilson coefficients  evaluated in ref.\cite{Neubert}
$$    c_{1}=1.105;\;\;  c_{2}=-0.245. $$
we get the theoretical values of the lifetimes of B-mesons which
are shown in Table.1:
\begin{eqnarray}
\begin{tabular}{||c|c|c|c|c||}\hline
&$\tau_{B_{q}}(ps)$ & $ \Gamma_{B_{q}}\times 10^{-13} {\rm GeV}$
& $\Gamma_{WE} / \Gamma_{b} $& $\Gamma_{PI} / \Gamma_{b}$  \\
\hline$ B^{0}$  & $1.56 $&$4.17 $ & $1.0 \%$ &  \\
 \hline $ B^{\pm} $ & $1.63 $&$4.05 $ &  & $-1.9 \%$  \\
\hline
\end{tabular}\nonumber
\end{eqnarray}
\vspace{0.5cm}

Table 1:  The lifetimes of $B^{0}$ and $B^{\pm}$ evaluated in our
model with the parameters obtained by fitting data of the
semi-leptonic decays, where
$\Gamma_{B_{q}}=\Gamma_{b}+\Gamma_{non}$.
\\

Thus theoretical result is $\tau(B^{-}) / \tau(B_{d}) = 1.03 $.
which is in general consistent with the data.\\

(b) For the heavy baryon $\Lambda_{b}$\\

The experiment data of $\Lambda_{b}$  are given \cite{Data}
\begin{eqnarray}
\left\{
\begin{array}{rl}
\tau_{\Lambda_{b}}\!\!\!\!&=(1.229 \pm 0.080) ps, \\
 BR(\Lambda_{b} \rightarrow \Lambda_{c} e \bar{\nu}_{e})\!\!\!\!&= (7.9 \pm
1.9)\%  ,\\
\Gamma_{SL}(\Lambda_{b} \rightarrow \Lambda_{c} e
\bar{\nu}_{e})\!\!\!\!&= (3.213 \sim  5.249) \times 10^{-14} {\rm
GeV}.
\end{array}
\right.
\end{eqnarray}
By fitting the data of the semi-leptonic decays of $\Lambda_b$, we
get the corresponding  parameters which are presented in Table 2.
Since the measured $\Gamma_{SL}$ which is the input parameter for
our model, has a relatively large tolerance range, the model
parameters can accordingly take various values, here we adopt four
typical values $\Gamma_{SL}-\sigma$, $\Gamma_{SL}$,
$\Gamma_{SL}+\sigma$ and $\Gamma_{SL}+2\sigma$, as input.
$\Gamma_{SL}$ and $\sigma$ are the central value and
standard deviation of the measured decay rate\cite{Data}.\\
\begin{table}[h]
\begin{eqnarray}
\!\!\!\!\!\!\!\!\!\!\!\!\!\!\!\!\!\!\!\!\!\!
\begin{tabular}{||c|c|c|c|c|c|c||}\hline
$\Gamma(\Lambda_{b} \rightarrow \Lambda_{c} e \bar{\nu}_{e})\times 10^{-14}
{\rm GeV} $ & $\alpha^{\Lambda_{b}}_{\rho}$ & $ \alpha^{\Lambda_{b}}_{\lambda}$
 & $\alpha^{\Lambda_{c}}_{\rho} $&$  \alpha^{\Lambda_{c}}_{\lambda} $
 &$\alpha^{\Lambda_{b}}_{Di}$&$\alpha^{\Lambda_{c}}_{Di}$ \\
\hline
\hline  3.213 (a)& $0.215$ & $0.601$ & $0.101$ & $ 0.364$ & $0.669$ &$0.435$ \\
 \hline
\hline  4.231 (b)& $0.221$ & $0.627$ & $0.114$ & $ 0.377$ & $0.712$ &$0.463$ \\
 \hline
 \hline 5.294 (c)& $0.234$ & $0.657$ & $0.127$ & $ 0.394$ & $0.745$ &$0.486$ \\
 \hline
 \hline  6.267 (d)& $0.255$ & $0.674$ & $0.142$ & $ 0.411$ & $0.788$ &$0.502$ \\
 \hline
\end{tabular}\nonumber
\end{eqnarray}
\vspace{0.5cm} Table 2:  These four sets of parameters which are
obtained by fitting data of semi-leptonic decay $\Lambda_{b}
\rightarrow \Lambda_{c} e \bar{\nu}_{e}$ and corresponding to (a)
($\Gamma_{SL}-\sigma$), (b) $\Gamma_{SL}$, (c)
($\Gamma_{SL}+\sigma$) and (d) ($\Gamma_{SL}+2\sigma$)
respectively.
\end{table}

\vspace{1cm}

With these parameters we can numerically calculate the lifetime of
$\Lambda_b$ similarly to what we have done for $B^0$ and
$B^{\pm}$. The results are shown in Table 3. \\

 \begin{eqnarray}
\begin{tabular}{||c|c|c|c|c||}
\hline
 Theory Values  &$\tau_{\Lambda_{b}}(ps)$ & $ \Gamma_{\Lambda_{b}}\times 10^{-13}
 {\rm Gev}$ & $\Gamma^{WA}_{non} / \Gamma_{b} $& $\Gamma^{PI}_{non} / \Gamma_{b}$  \\
\hline
\hline  3-valence-quark model (a) &$1.50$ & $4.39 $ &  $6.4 \%$ &  less than $ 1 \%$ \\
\hline   diquark Model \ \ \ \ \ \ \ \ \ \ \ \ (a)& $1.50$ &$4.36 $  &  $ 5.6 \%$  & less than $ 1 \%$  \\
\hline
\hline  3-valence-quark model  (b)&$1.48 $ & $4.44 $ &  $7.5 \%$ &  less than $ 1 \%$ \\
\hline  diquark Model \ \ \ \ \ \ \ \ \ \ \ \ (b) & $1.50$ &$4.40 $  &  $ 6.6 \%$  & less than $ 1 \%$  \\
\hline
\hline  3-valence-quark model  (c)&$1.46 $ & $4.50 $ &  $9.0 \%$ &  less than $ 1 \%$ \\
\hline  diquark Model  \ \ \ \ \ \ \ \ \ \ \ \ (c)& $1.48$ &$4.44 $  &  $ 7.4 \%$  & less than $ 1 \%$  \\
\hline
\hline  3-valence-quark model  (d)&$1.43 $ & $4.60 $ &  $11.3 \%$ &  less than $ 1 \%$ \\
\hline  diquark Model  \ \ \ \ \ \ \ \ \ \ \ \ (d)& $1.47$ &$4.49 $  &  $ 8.6 \%$  & less than $ 1 \%$  \\
\hline
\end{tabular}\nonumber
\end{eqnarray}
\vspace{0.5cm}

Table 3: The lifetime of $\Lambda_{b}$ evaluated in our model with
different sets of the parameters obtained by fitting data of the
semi-leptonic decays, and
$\Gamma_{\Lambda_{b}}=\Gamma_{b}+\Gamma_{non}$.\\

The theoretical results indicate that $ \tau(\Lambda_{b}) /
\tau(B_{d})\sim 0.91\sim 0.95$ in the 3-valence quark model and $
\tau(\Lambda_{b}) / \tau(B_{d})\sim 0.93\sim 0.96$ in the diquark
model while the measured value of the ratio is 0.79.

\section{Conclusion and Discussion}
The Standard Model (SM) is definitely responsible for the weak
transition and the lifetimes of B-mesons, as well as $\Lambda_{b}$
because they do not have strong decay channels. By the common
understanding, such inclusive decay modes are dominated by the
spectator mechanism, namely the decay rate is almost fully
determined by the decays of the heavy flavor in the hadron.
However, some puzzles in b-physics emerge as indicated by many
authors \cite{Lenz}, such as the lifetime of $\Lambda_{b}$, charm
number missing in B  decays etc. Among them the lifetime of
$\Lambda_{b}$ raises the most challenging problem.

Bigi et al. discussed the lifetime difference of  $D^{\pm}$  and
$D^{0}$ \cite{Bigi}, they showed that the non-spectator effects
play a crucial role  to the lifetimes. In fact, the Pauli
interference(PI) mechanism greatly suppresses the decay rate of
$D^{\pm}$ compared to $D^{0}$, and  QCD  can almost perfectly
explain why  $\tau(D^{\pm}) \sim
 2.55 \tau(D^{0})$, but  $\tau(B^{\pm}) \sim
  \tau(B^{0})$. Therefore it is natural to consider to
take into account the non-spectator effects for evaluating the
lifetime of $\Lambda_{b}$. Even though as Bigi et al. indicated,
such effects are not so important for b-physics, one may still
think that the contributions may be not negligible, because the
baryon structure is different from that of mesons, at least there
are two light valence quarks and each of them can join the b-quark
to make a non-spectator contribution. Moreover, there is a diagram
where three valence quarks(b u d) participate in the reaction (see
Fig1(c)), and it does not appear for the meson case. Unfortunately
our numerical results indicate that this contribution is too small
to result in any substantial change to the lifetime of
$\Lambda_{b}$.

The derivation at quark level is standard, the main difficult part
comes from the evaluation of the hadronic matrix elements. We
employ the simplest non-relativistic harmonic oscillator model
which has been proved to be successful in phenomenology, but
definitely is not accurate. We determine the concerned parameters
by fitting the data of semi-leptonic decays and then use them to
evaluate the lifetimes. In the process we can reduce the errors
and uncertainties in the model-dependent calculation. Then we
repeat the procedure with the same model to deal with the B
mesons, thus when we compute the ratios of $\tau(B^{\pm}) /
\tau(B^{0})$ and $\tau(\Lambda_{b}) / \tau(B^{0})$, the
model-dependence is further reduced.

For $\Lambda_{b}$, we employ two pictures, the three-valence-quark
picture and one-heavy-quark-one-light-diquark picture to evaluate
the lifetime of $\Lambda_{b}$, the result obtained in the two
pictures are qualitatively consistent.

Our numerical results indicate that the non-spectator effects can
only result in about at most $7\%$ reduction of the lifetime of
$\Lambda_{b}$ whereas  the experimental data demand a $21\%$
reduction. Therefore we can claim that the non-spectator
contribution is sizable and cannot be neglected, taking into
account the effects can remarkably alleviate the discrepancy which
was not improved by just changing the matrix element parameters
\cite{Liu}. However, on the other side, the discrepancy still
stands and demands a more satisfactory explanation. Because the
large fraction of $21\%$ deviation in the ratios cannot be
compensated by only the non-spectator effects, there must be some
unknown mechanisms which induce the difference between baryon and
meson. For example, the proposed three-body force in baryons
\cite{Isgur} might lead to a dramatic difference between baryon
and meson, all these mechanisms are worth careful studies and may
provide us with better understanding of the hadron structure and
fundamental interactions.

As the verbal work of the paper is near completion, we notice
Gabbiani et al's work\cite{Gabbiani}, they also evaluate the
non-spectator effects on the lifetime of $\Lambda_{b}$, even
though in a different approach, our results
are consistent with theirs.\\

\noindent{\bf Acknowledgement}\\
 This work is partially supported by the National
Natural Science Foundation of China. Our preliminary results of
this work was presented at the conference organized by the NNSF of
China.\cite{pro}.

\noindent{\bf Appendix A}\\
 The semileptonic and non-leptonic
decay rates of b-quark up to order $1/m^{2}_{Q}$ are given as
following\cite{Voloshin}.
\begin{eqnarray}
\Gamma_{SL}(H_{b})&=&\Gamma^{(b)}_{0} \eta(x_{c},x_{l},0)
[I_{0}(x_{c},0,0) < H_{b} \mid \bar{b} b \mid H_{b} > -\frac{2 <
\mu^{2}_{G}
>_{H_{b}}}{m^{2}_{b}} I_{1}(x_{c},0,0)  ] \nonumber \\
\Gamma_{NL}(H_{b})&=&\Gamma^{(b)}_{0} N_{c}
\{(c^{2}_{1}+c^{2}_{2}+ \frac{2 c_{1} c_{2}}{N_{c}}) [(\alpha
I_{0}(x_{c},0,0) +\beta I_{0}(x_{c},x_{c},0))< H_{b} \mid \bar{b}
b \mid H_{b} > \nonumber \\
 &-& \frac{2 < \mu^{2}_{G}
>_{H_{b}}}{m^{2}_{b}} ( I_{1}(x_{c},0,0) +I_{1}(x_{c},x_{c},0) )  ]\nonumber \\& -& \frac{8 < \mu^{2}_{G}
>_{H_{b}}}{m^{2}_{b}} \frac{2 c_{1} c_{2}}{N_{c}} ( I_{2}(x_{c},0,0) +I_{2}(x_{c},x_{c},0) )  ]\},
\end{eqnarray}\\
where\\
\begin{equation}
\Gamma^{(b)}_{0}=\mid V_{cb} \mid^{2} \frac{G_{F}^{2}
m^{5}_{b}}{192 \pi^{3}},
\end{equation}\\
and the following notations have been adopted : $I_{0}$,$I_{1}$
and $I_{2}$ are the phase-space factors, namely
\begin{eqnarray}
I_{0}(x,0,0) &=& (1-x^{2})(1-8x+x^{2})-12 x^{2} \log x,\nonumber \\
I_{1}(x,0,0) &=& \frac{1}{2}(2-x \frac{d}{dx})I_{0}(x,0,0),\nonumber \\
I_{2}(x,0,0) &=& (1-x)^{3},\nonumber \\
I_{0}(x,x,0) &=& v (1-14x-2x^{2}-12x^{3})+24 x^{2}(1-x^{2}) \log \frac{1+v}{1-v},\nonumber \\
I_{1}(x,x,0) &=& \frac{1}{2}(2-x \frac{d}{dx})I_{0}(x,x,0),\nonumber \\
I_{2}(x,x,0) &=& v(1+\frac{x}{2}+ 3x^{2} )-3 x(1-2x^{2}) \log
\frac{1+v}{1-v},\nonumber \\
x_{c}&=& \frac{\bar{m}^{2}_{c}}{m^{2}_{b}},v=\sqrt{1-4x},
\end{eqnarray}\\
with $I_{0,1,2}(x,x,0)$ describing the $b \rightarrow c \bar{c} s$
transitions.  $\eta(x_{c},x_{l},0)$ is the QCD radiative
correction to the semileptonic decay rate  and its general
analytic expression is given in \cite{Hokim}. In a special case
the expression of $\eta(x,0,0)$ is given in \cite{Nir} and
numerically it can be approximated as \cite{C.S.Kim,M.Luke}.
\begin{equation}
\eta(x,0,0)  \cong 1 - \frac{ 2 \alpha_{s} }{3 \pi} [
(\pi^{2}-\frac{31}{4})(1-\sqrt{x})^{2} +\frac{3}{2}].
\end{equation}\\
For the decay $b \rightarrow c  \tau \bar{\nu} $, according to
ref.\cite{Bagan} we roughly have \\
\begin{equation}
\Gamma(b \rightarrow c \tau \bar{\nu}_{\tau} ) \sim 0.25 \Gamma_{b
\rightarrow e \bar{\nu}_{e}}.
\end{equation}
For the dimension-three operator $\bar{Q}Q$, the expectation value
can be expressed as
\begin{equation}
< H_{Q} \mid \bar{Q} Q \mid H_{Q} > =1-\frac{ < (p_{Q})^{2}
>_{H_{Q}}}{2 m^{2}_{Q}}+\frac{ < \mu^{2}_{G}
>_{H_{Q}}}{ 2 m^{2}_{Q}} +\mathcal{O}(1/m^{3}_{Q}),
\end{equation}\\
where $< (p_{Q})^{2}
> \equiv < H_{Q} \mid \bar{Q} (i D)^{2}Q \mid H_{Q} >  $ denotes
the average kinetic energy of the quark Q moving inside the hadron
and $< \mu^{2}_{G}
 > \equiv < H_{Q} \mid \bar{Q} \frac{i}{2} \sigma  \cdot G Q \mid H_{Q}
 >$.\\
According to ref.\cite{Voloshin} the kinetic energies of b and c
quarks are respectively
\begin{equation}
\frac{ < (p_{b})^{2}
>_{B}}{2 m^{2}_{b}} \simeq 0.016.
\end{equation}
For the QCD operator, one finds $< \mu^{2}_{G}
 >_{P_{Q}} \equiv \frac{3}{2} m_{Q} (M_{V_{Q}}-M_{P_{Q}})$, where
 $P_{Q}$ and $V_{Q}$ denote the psendoscalar and vector mesons,
respectively.

\begin{figure}[b] \centering  \scalebox{1.1}{
\includegraphics{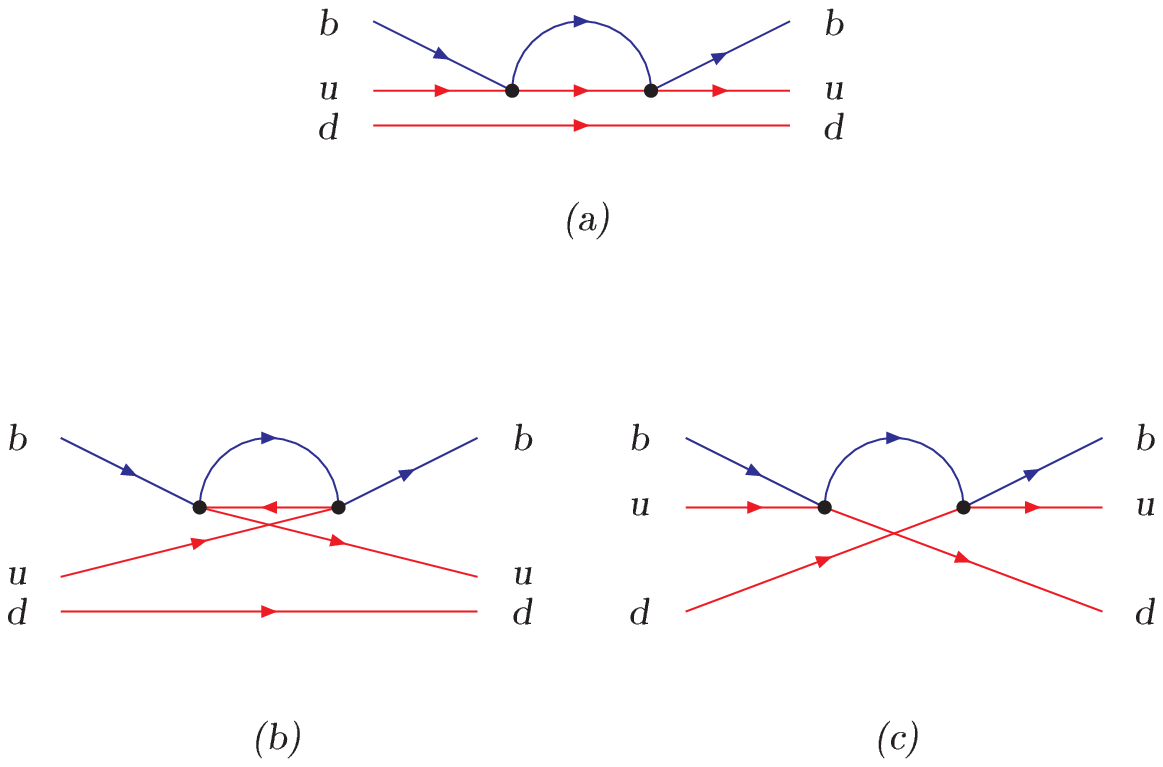}} \caption{The Feynman diagrams for the
non-spectator effects in the 3-valence-quark-picture picture, (a)
is the WE and (b) is the PI  where only one of the light quarks
exchanges W-boson with the b-quark while the rest one stands by as
a spectator. The Pauli interference is only a terminology and its
real meaning is explained in the footnote of the text. Fig.1 (c)
refers to a "PI" and "WE" mixed Feynman diagram where all the
three quarks take part in the reaction.} \label{Fig.1}
\end{figure}

\begin{figure}[b] \centering  \scalebox{1.1}{
\includegraphics{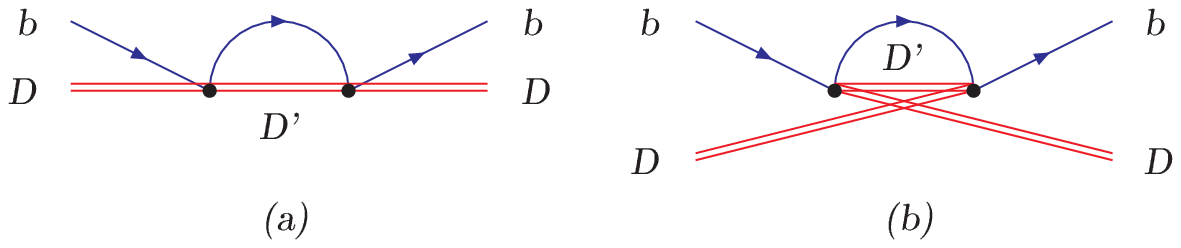}} \caption{The Feynman diagrams for the
non-spectator effects in the one-heavy-quark-one-light-diquark
picture, (a) is the WE, and (b) is PI   where  the light diquark
exchanges W-boson with the b-quark. The Pauli interference is only
a terminology. } \label{Fig.2}
\end{figure}


\begin{thebibliography}{99}
\bibitem{Data}The Data Group, Phys. Rev. \textbf{D66}(2002)010001;
\bibitem{Isgur}N. Isgur and M. Wise, Phys. Lett.
 \textbf{B232}(1989)113;  \textbf{B237}(1990)527;
\bibitem{Georgi}H. Georgi, Nucl. Phys. \textbf{B361}(1991)339;  M. Neubert, Phys. Rep. \textbf{245}(1994)259;
\bibitem{Beneke}M. Beneke, G. Buchulla, M. Neubert and C. Sachrajda, Phys. Rev. Lett.  \textbf{83}(1999)1914; Nucl. Phys. \textbf{B591}(2000)313;
\bibitem{Bigi}I. Bigi, N.
Uraltsev Phys. Lett. \textbf{B280}(1992)120; I. Bigi, N. Uraltsev
and A. Vainshtein, Phys. Lett. \textbf{B293}(1992)430; (E)
\textbf{B297}(1993)477;  B. Blok,  M. Shifman, Nucl. Phys.
\textbf{B399}(1993)441; 459; I. Bigi, B. Blok, M. Shifman, N.
Uraltsev and A. Vainshtein, \textit{In B Decays}, edited by S.
Stone, World Scientific Pub.Co.Singapore 1994;  G. Belliui et al,
Phys. Rep. \textbf{289}(1997)1;  M. Neubert and C. Sachrajda,
hep-ph/9603202; W. Dai et al, Phys. Rev. \textbf{D60}(1999)034005;
\bibitem{He}N. Deshpande and X. He,  Phys. Lett. \textbf{B336}(1994)471;
\bibitem{N.Bilic}N. Bili$\acute{c}$, B. Guberina, J. Trampeti$\acute{c}$, Nucl. Phys.
\textbf{B248}(1984)261;  M. Voloshin, M. Shifman, it Yad. Fiz.
\textbf{41}(1985)187 [\textit{Sov. Journ. Nucl. Phys.
\textbf{41}(1985)120}]; \textit{ZhETF} \textbf{91}(1986)1180
[\textit{Sov.  Phys. - JETP \textbf{64}(1986)968}];
\bibitem{Mohanta}R. Mohanta, A. K. Giri, M. P. Khanna, M. Ishida, S. Ishida and   M. Oda, hep-ph/9908291; N. Isgur, D. Scora, B. Grinstein and M. Wise,
Phys. Rev. \textbf{D39}(1989)799;
\bibitem{D. Chakraverty}D. Chakraverty, T. De and B.
Datta-Roy, hep-ph/9612369;
\bibitem{Olivier}A. Le Yaouanc, L. Olivier, O. P$\grave{e}$ne and J. -C. Raynal,  \textit{Hadron transitions in the quark model}, Gordon and Breach Science Publish(1988);
\bibitem{Bhaduri}R. K. Bhaduri, \textit{Models of The Nucleon From Quarks to
Solition}, AddisonWesley(1988)
\bibitem{Bagan}E. Bagan, P. Ball, V. Braun and
P. Gasdzinsky, Nucl. Phys. \textbf{B342}(1995)362; erradum, ibid
\textbf{B372}(1996)363;
\bibitem{Neubert}M. Neubert and C. Sachrojda, Nucl. Phys.
\textbf{B483}(1997)339.
\bibitem{Hou}W-S. Hou and B. Tseng, Phys. Rev. Lett. \textbf{80}(1998)343.
\bibitem{Dosch}H. G. Dosch, E. M. Ferreira and  F. S. Navarra, M.
Nielsen, Phys. Rev. \textbf{D65} (2002) 114002;
\bibitem{Liu}Chao-Shang Huang, Chun Liu and Shi-Lin Zhu, Phys. Rev. \textbf{D65} (2000) 054004;
\bibitem{Lenz} A. Lenz, U. Nierste and G. Ostemaier, Phys. Rev.
{\bf D56} (1997) 7228; A. Lenz, hep-ph/0010099.
\bibitem{Gabbiani} F. Gabbiani, A. Onishchenko and A. petrov,
hep-ph/0303235.
\bibitem{Voloshin}M. Voloshin and M. Shifman, Sov. J. Nucl. Phys. \textbf{41}(1985)120;
I. Bigi, M. Shifman, N. Uraltsev and A. Vainshtein, Phys. Rev.
Lett. \textbf{71}(1993)496; B. Blok and M. shifman, Nucl. Phys.
\textbf{B399}(1993)459; A. Manohar and M. Wise, Phys. Rev.
\textbf{D49}(1994)1310; I. Bigi, Phys. Lett.
\textbf{B371}(1996)105; H. Y. Cheng, Phys. Rev.
\textbf{D56}(1997)2783.
\bibitem{Hokim}Q. Hokim and X. Y. Pham, Phys. Lett. \textbf{112B}(1983)297.
\bibitem{Nir}Y. Nir, Phys. Lett. \textbf{B221}(1989)184.
\bibitem{C.S.Kim}C. S. Kim and A. D. Martin, Phys. Lett. \textbf{B225}(1989)186.
\bibitem{M.Luke}M. Luke, M. J. savage, M. B. Wise, Phys. Lett. \textbf{B345}(1995)301.
\bibitem{pro} Talk presented at the NNSF report conference,
 Jan. 2003, Beijing, China and will be published in the
proceddings of the conference, edited by the NNSF of China.
\end{thebibliography}
\end{document}